\documentstyle[12pt,aps,prd]{revtex}
\begin{document}
\tightenlines
\vspace{.5cm}
\begin{center} {\large\bf On the n=4 Supersymmetry for the FRW Model}
\vspace{.5cm}\\
A. Pashnev${}^a$\footnote{e-mail: pashnev@thsun1.jinr.ru},
J.J. Rosales${}^{b}$\footnote{e-mail:
juan@ifug3.ugto.mx}, V.I. Tkach ${}^b$\footnote{e-mail:
vladimir@ifug3.ugto.mx}, and M. Tsulaia ${}^a$\footnote{e-mail:
tsulaia@thsun1.jinr.ru}

 \vspace{.5cm}
 ${}^a${\it Bogoliubov Laboratory of Theoretical Physics, JINR} \\
 {\it Dubna, 141980, Russia}\\
 ${}^b${\it Instituto de Fisica,
 Universidad de Guanajuato\\
 C.Postal 66318, 05315-970 Leon, Gto. Mexico}\vspace{.5cm}\\
 {\bf Abstract}
 \end{center}
 In this work we have constructed the $n=4$ extended local conformal time
supersymmetry for the Friedmann-Robertson-Walker (FRW) cosmological models.
This is based on the superfield construction of the action, which is invariant
under worldline local $n=4$ supersymmetry with
$SU{(2)}_{local}\otimes SU{(2)}_{global}$ internal symmetry. It is shown
that the supersymmetric action has the form of the localized
(or superconformal) version of the action for $n=4$ supersymmetric quantum
mechanics. This superfield procedure provides a well defined scheme for
including supermatter.

\vspace{0.2cm}

 \newpage
 \section{Introduction}
The supersymmetric quantum mechanics (SUSY QM), being introduced first in
\cite{1} for the $n=2$ case turns out to be convenient tool for
investigating problems of the supersymmetric field theories, since it
provides the simple and at the same time quite adequate understanding of
various phenomena arising in the relativistic theories. For instance,
one-dimensional \cite{2} and multidimensional \cite{3,4} $n=4$ (SUSY QM)
can be associated with $n=1$, $D=4$ supersymmetric field theories
(including supergravity) subject to an appropriate dimensional reduction
down to $D=1$. Applications of supersymmetric quantum mechanics to the theory
of black holes and to the other problems have been reviewed in \cite{5,6}. This
underlines the importance of studying simplified models in order to
understand the key features of more complicated four dimensional problems.
Along this research line, in recent years a systematic way to construct
models of local supersymmetric quantum cosmology has been proposed
\cite{7,8}. This was performed by introducing a superfield formulation. This
is because superfields defined on superspace allow all the component fields
in a supermultiplet to be manipulated simultaneously in a manner wich
automatically preserves supersymmetry. This approach has the advantage
of being simpler than the proposed models based on full supergravity
\cite{9,10}, and by means of this local symmetry procedure it gives the
corresponfing fermionic partners in a direct manner. Using the superfield
formalism the $n=2$ local supersymmetric quantum mechanics for the
cosmological models was constructed in \cite{7,8}, and a normalizable
wavefunction of the Universe for the FRW cosmological
model was obtained in \cite{11}. Although these
models do not attempt to describe the real world, they keep many of the
features occurring in four dimensional space-time, which could really be
studied in the quantum version of simplified models.

The most physically interesting case is provided by the $n=4$ local
supersymmetry, since it can be applied to the description of the systems
resulting from the ``realistic'' $n=1, D=4$ supergravity subject
to an appropriate dimensional reduction down to $D=1$.

In the present report we propose a description of the FRW model based on the
superfield construction of an action invariant under $n=4$ worldline local
supersymmetry with the $SU{(2)}_{local}\otimes SU{(2)}_{global}$ internal
symmetry. Due to the invariance of the action we obtain the constraints,
which form a closed superalgebra of the $n=4$ supersymmetric quantum mechanics.

 \section{ $n=4$ superconformal transformations}

 We begin by considering the homogeneous and isotropic metric defined by

 \begin{equation}
 ds^2 = - N^2(t)dt^2 + R^2(t)d\Omega^2_3,
 \label{1}
 \end{equation}
 where $d\Omega^2_3$ is the spatial FRW
 standard metric over three-space. The lapse function $N(t)$ and the scale
 factor $R(t)$ depend on the time parameter only. To construct the
 superfield action in the worldline superspace $(t,\theta^a, \bar\theta_a)$
 (with $t$ being a time
 parameter, and $\theta^a$ and $\bar\theta_a=(\theta^a)^*,
 ~(a=1,2 $ is  an SU(2) index) being two complex Grassmann coordinates) one
introduces a real ``matter'' superfield
 ${I\!\!R}(t,\theta^a, \bar\theta_a)$ and a
 worldline supereinbein ${I\!\!N}(t,\theta^a, \bar\theta_a)$ which have
 the following properties with respect to the $SU(2)$ $n=4$
 superconformal transformations of the worldline superspace \cite{12}
 \footnote{Our conventions for spinors are as follows:
 ${\theta}_{a} ={\theta}^{b}{\varepsilon}_{ba},\; {\theta}^{a}=
 {\varepsilon}^{ab}{\theta}_{b},\; {\bar \theta}_{a}={\bar
 \theta}^{b}{\varepsilon}_{ba},\; {\bar \theta}^{a}=
 {\varepsilon}^{ab}{\bar \theta}_{b},\; {\bar \theta}_a =
 (\theta^a)^\ast,\; {\bar \theta}^a = -(\theta_a)^\ast,\; (\theta
 \theta)\equiv \theta^a \theta_a = -2 \theta^1 \theta^2,\;
  (\bar \theta \bar \theta)\equiv \bar \theta_a \bar \theta^a = (\theta
 \theta)^\ast,\;  (\bar \theta \theta) \equiv \bar \theta_a \theta^a,\;
  \varepsilon^{12}=- \varepsilon^{21}= 1,\;
 \varepsilon_{12} = 1$.}

 \begin{eqnarray}
 \delta t &=& \Lambda-\frac{1}{2}\theta^aD_a\Lambda - \frac{1}{2}
 \overline{\theta}_a\overline{D}^a\Lambda,\qquad
 \delta\theta^a = i\overline{D}^a\Lambda,\qquad
 \delta\overline{\theta}_a=iD_a\Lambda,
 \label{2}
 \end{eqnarray}
 \begin{equation}
 \delta {I\!\!R} = -\Lambda\dot {I\!\!R}+ \dot\Lambda{I\!\!R}-
i(D_a\Lambda)
 (\overline{D}^a{I\!\!R})-i(\overline{D}^a\Lambda)( D_a{I\!\!R}),
 \label{3}
 \end{equation}
 \begin{equation}
 \delta {I\!\!N} = -\Lambda\dot{{I\!\!N}}-\dot\Lambda {I\!\!N}-
 i(D_a\Lambda)(\overline{D}^a {I\!\!N})-i(\overline{D}^a\Lambda)
 (D_a {I\!\!N})
 \label{4},
 \end{equation}
 where dot denotes the time derivative $\frac{d}{dt}$. The
 transformation law (\ref{3}) for the superfield ${I\!\!R}$ shows that
this superfield is a vector superfield in the one-dimensional
$n=4$ superspace,
while the superfield ${I\!\!N}{I\!\!R}$ is a scalar \cite{12}.

The superfields ${I\!\!R}$ and ${I\!\!N}$ obey the quadratic constraints
 \begin{eqnarray}
 \lbrack D_a , \overline {D}^a\rbrack {I\!\!R} &=& -4m, \qquad\qquad
 D^a D_a {I\!\!R} = 0, \qquad\qquad
 \overline{D}_a \overline{D}^a {I\!\!R} = 0,
 \label{5}
 \end{eqnarray}
 and
 \begin{eqnarray}
 \lbrack D_a , \overline{D}^a\rbrack \frac{1}{{I\!\!N}} &=& 0, \qquad
 D^a D_a \frac{1}{{I\!\!N}} = 0,
 \qquad \overline {D}_a \overline {D}^a \frac{1}{{I\!\!N}} =0,
 \label{6}
 \end{eqnarray}
 where
 \begin{equation}
 D_a = \frac{\partial}{\partial \theta^a}- \frac{i}{2} \overline{\theta}_a
 \frac{\partial}{\partial t} ,
 \quad
 \overline{D}^a = \frac{\partial}{\partial \overline{\theta}_a}
 - \frac{i}{2} \theta^a \frac{\partial}{\partial t},
 \label{7}
 \end{equation}
 are the supercovariant derivatives, and $m$ is an arbitrary constant. The
 infinitesimal superfield
 \begin{eqnarray}
 \Lambda(t,\theta,\overline{\theta})&=& a(t)
 +\theta^a \overline{\alpha}_a(t)
 -\overline{\theta}_a \alpha^a(t)
 + \theta^a(\sigma^i)_a^b \overline{ \theta}_b b_i(t)
 \nonumber\\
 &&+\frac{i}{4}(\theta \theta) \overline{\theta}_a
\dot{\overline \alpha}^a(t) -
 \frac{i}{4}(\overline{\theta} \overline{\theta}) \theta^a {\dot \alpha}_a(t)
 +\frac{1}{16}(\theta\theta)(\overline{\theta} \overline{\theta}) \ddot{a}(t),
 \label{8}
 \end{eqnarray}
 contains the parameters of local time reparametrizations $a(t)$,
 local supertranslations $\alpha(t)$, $\overline \alpha(t)$. $b_i(t)$
 being a local SU(2) parameter of the worldline superspace.
 The constraint (\ref{5}) can be explicitly solved,
 the solution being described by the superfield
 \begin{eqnarray}
 {I\!\!R}(t,\theta,\overline{\theta})&=& R^{\prime}(t)
 +\theta^a {\overline \lambda}_a'(t)
 -{\overline \theta}_a \lambda'~^a(t)+
 \theta^a(\sigma_i)_a~^b {\overline \theta}_b F'_i(t)
 + m(\theta \overline{\theta}) \label{9}\\
 &&+\frac{i}{4}(\theta \theta){\overline \theta}_a
\dot{\overline \lambda}^{\prime}~^a
 - \frac{i}{4}(\overline{\theta} \overline{\theta})
\theta^a {\dot \lambda}^{\prime}_a
 +\frac{1}{16}(\theta\theta)(\overline{\theta} \overline{\theta})
\ddot{R^{\prime}}(t).\nonumber
\end{eqnarray}
This superfield contains one bosonic field $R'$, which is associated
with the scale factor $R(t)$ (see Eq. ({\ref12})) of the FRW model, the
Grassmann-odd (they are  four) fermionic fields $\lambda^a(t)$ and
${\bar\lambda}_a(t)$ are their superpartners
being spin degrees of freedom, and
 $F_a^b = {(\sigma^i)_a^b} F_i$ are three auxiliary fields
 where $(\sigma^i)_a^b$ (i=1,2,3) are the ordinary Pauli matrices.

 The constraint (\ref{6}) being described by the superfield
 \begin{eqnarray}
 \frac{1}{{I\!\!N}}(t,\theta,\overline{\theta})&=&\frac{1}{N(t)}
 +\theta^a \overline{\psi}^{\prime}_a(t)
 -\overline{\theta}_a \psi^{\prime}~^a(t)
 +\theta^a (\sigma^i)_a~^b {\overline \theta}_b
 V'_i(t) \label{10}\\
 && +\frac{i}{4} (\theta \theta)\overline{\theta}_a
\dot{\overline \psi}^{\prime}~^a(t)
 - \frac{i}{4} (\bar\theta \bar\theta)\theta^a \dot{\psi}^{\prime}_a
 + \frac{1}{16}(\theta\theta)(\bar\theta \bar\theta)
 \frac{d^2}{dt^2}\frac{1}{N(t)}.\nonumber
 \end{eqnarray}
 The superfield ${I\!\!N}$ describes an $n=4$
 worldline supergravity multiplet consisting of the einbein
 ``graviton" $N(t)$, two complex ``gravitinos" $\psi^{\prime a}(t)$
 and ${\overline\psi}^{\prime}_a(t)$, and the $SU(2)$ gauge field
 $V^{\prime}_i(t)$. The components of ${I\!\!N}$ play the role of Lagrange
multipliers. Their presence implies that the dynamics of the model is subject
to constraints.

The $n=4$ superfield action for the FRW model invariant under the $n=4$
superconformal symmetry has the following form \cite{12}

 \begin{equation}
 S= \frac{8}{{\mbox{\ae}^2}} \int dt d^2\theta d^2 \overline{\theta}
 {I\!\!N}^2 {I\!\!R}^3,
 \label{11}
 \end{equation}
 where ${\mbox{\ae}}^2 = 8\pi G_N$, $G_N$ is the Newtonian constant of gravity.
 So, integrating (\ref{11}) over the Grassmann coordinates $\theta$,
$\overline\theta$ and making the following redefinition of the
 component fields

\begin{eqnarray}
 \psi &=& N^{\frac{3}{2}}\psi^{\prime}, \quad
 V_i = 2N(V^{\prime}_i + N (\psi^{\prime}\sigma_i \overline{\psi}^{\prime})),
  \quad
 \lambda = {\sqrt N}(\lambda^{\prime} - R \psi^{\prime}), \label{12}\\
 F_i &=& 2{\sqrt N}\lbrace F^{\prime}_i - R V^{\prime}_i +
\frac{\sqrt N}{2}(\psi^{\prime} \sigma_i \overline \lambda) +
\frac{\sqrt N}{2}(\lambda\sigma_i {\overline \psi}^{\prime})\rbrace
\qquad
 R = NR^{\prime}, \nonumber
 \end{eqnarray}
 one obtains the component action

 \begin{eqnarray}
 S &=&\frac{1}{{\mbox{\ae}}^2} \int \Big\{ -\frac{3R (D R)^2}{ N} -
 {6iR}\lbrack (\overline{\lambda} D\lambda) +
(\lambda D\overline{\lambda})\rbrack
 - {3R} F_i F^i + 6 \sqrt N
(\lambda \sigma^i \overline{\lambda}) F_i \label{13}\\
&& - 3 \lbrack (\overline{\lambda}
\overline{\lambda})
(\lambda \psi) + (\overline{\lambda} \overline{\psi})
(\lambda \lambda)\rbrack +  12 m^2 NR
 + 12 mN (\overline{\lambda} \lambda) -
 12 mR \lbrack (\overline{\lambda} \psi) +
(\overline{\psi} \lambda)\rbrack \Big\}dt,
\nonumber
 \end{eqnarray}
where $DR = \dot R -i\lbrack (\overline{\psi} \lambda) -
(\overline{\lambda} \psi)\rbrack $ is the
supercovariant derivative, $D\lambda = {\dot \lambda} +
\frac{i}{2}(\sigma^i) \lambda V_i$ and its conjugate are the SU(2)
covariant derivatives. Upon solving for the equations of motion of the
auxiliary fields $F_i$, substituting the solution back into (\ref{13}) and
making the redefinitions
 $\lambda \to {\mbox{\ae}}  (6R)^{-1/2} \lambda$,
 $\overline{\lambda} \to {\mbox{\ae}} (6R)^{-1/2} \overline{\lambda}$, and putting
$4m^2 = k$ (where  $k = 1, 0, -1$ stands for spherical, plane or
hyperspherical three space), we have then

 \begin{eqnarray}
 S &=& \int \Big\{- \frac{3R (D R)^2}{{\mbox{\ae}}^2 N} -
 i\lbrack (\overline{\lambda} D\lambda) +
(\lambda D\overline{\lambda})\rbrack +
 \frac{3 k}{{\mbox{\ae}}^2} NR + \frac{N \sqrt{k}}{R} (\overline{\lambda} \lambda)
\label{14}\\
&& - \frac{\sqrt{ k} \sqrt{6R}}{{\mbox{\ae}}} \lbrack (\overline{\lambda} \psi) +
(\overline{\psi} \lambda)\rbrack
 - \frac{{\mbox{\ae}}}{\sqrt{24} R^{3/2}}\lbrack
(\overline{\lambda} \overline{\lambda})
(\lambda \psi) + (\overline{\lambda} \overline{\psi})
(\lambda \lambda)\rbrack -
\frac{N{\mbox{\ae}}^2}{8R^3}(\overline{\lambda} \overline{\lambda})
(\lambda \lambda) \Big\} dt.
\nonumber
\end{eqnarray}
The expression (\ref{13}), as well as the expressions
(\ref{17})-(\ref{19}) for the Hamiltonian and Supercharges, can be considered
also for the hyperspherical geometry at the value $4m^2 = -1$ when they
lost their hermiticity properties. At least formally this should not lead to
contradictions in the physical sector which is singled out by the equations
(\ref{32}) in the quantum case.

 Performing Legendre transformations one arrives at the first order form
 of the action

 \begin{equation}
 S= \int \Big\{P \dot R + i\lbrack (\lambda \dot{\overline{\lambda}})
 + (\overline {\lambda} \dot{\lambda})\rbrack - H_c \Big\} dt,
 \label{15}
 \end{equation}
 where $P$ is the momentum canonically conjugate to $R(t)$, and
 the canonical Hamiltonian $H_c$ of the system has the following structure
 \begin{equation}
 H_c = N(t) H_{0} + i\psi^a(t) \overline {Q}_a +
 i\overline {\psi}_a(t) Q^a - V^i(t) {\cal F}_i,
 \label{16}
 \end{equation}
 with
 \begin{equation}
 H_{0}= -\frac{{\mbox{\ae}}^2 P^2}{12R} -
\frac{3kR}{{\mbox{\ae}}^2}
 - \frac{\sqrt {k}}{R} (\overline{\lambda} \lambda) +
 \frac{{\mbox{\ae}}^2}{8R^3} (\overline{\lambda} \overline{\lambda})
(\lambda \lambda)
\label{17}
 \end{equation}

 \begin{equation}
 {\overline{Q}}_a = \Big(\frac{{\mbox{\ae}} R^{-1/2} P}{\sqrt {6}} +
 i \frac{\sqrt{k} \sqrt{6R}}{{\mbox{\ae}}} \Big ) {\overline{\lambda}_a} -
 3i{\mbox{\ae}} (6 R)^{-3/2} ({\overline{\lambda}} {\overline{\lambda}})
\lambda_a,
 \label{18}
 \end{equation}

 \begin{equation}
 {Q}^b = \Big(\frac{{\mbox{\ae}} R^{-1/2} P}{\sqrt {6}} -
 i \frac{\sqrt{k} \sqrt{6R}}{{\mbox{\ae}}} \Big) \lambda^b -
 3i{\mbox{\ae}} (6R)^{-3/2} {\overline \lambda}^b (\lambda \lambda),
 \label{19}
 \end{equation}

 and
\begin{equation}
{\cal F}_i = (\lambda \sigma_i \overline{\lambda}),
\label{20}
\end{equation}
where $H_0$ is the Hamiltonian of the system, $Q^a$ and $\overline{Q}_a$ are
the supercharges, and ${\cal F}_i$ are the SU(2) rotations.
These formulae for the conserved supercharges complete the classical
description of the desired $n=4$ SUSY QM and now to quantize it we should
analyze it's constraints. Following the standard procedure of quantization
of the system with bosonic and fermionic degrees of freedom, we
introduce the canonical Poisson brackets

\begin{eqnarray}
\lbrace R,P \rbrace &=& 1, \qquad
\lbrace \lambda^a, \pi_{(\lambda)_b} \rbrace = - \delta^a_b, \qquad
\lbrace \overline{\lambda}_a, \pi^b_{(\overline{\lambda})} \rbrace =
-\delta_a^b,
\label{21}
\end{eqnarray}
where $P, \pi_{(\lambda)}$ and $\pi_{(\overline{\lambda})}$ are the
momenta conjugated to $R, \lambda^a$ and $\overline{\lambda}_a$. From the
explicit form of the momenta
\begin{eqnarray}
P &=& -\frac{6R}{{\mbox{\ae}}^2 N}DR =
-\frac{6R}{{\mbox{\ae}}^2 N}\lbrace {\dot R} -
\frac{i {\mbox{\ae}}}{\sqrt{6R}}\lbrack (\overline{\psi} \lambda) -
(\overline{\lambda} \psi)\rbrack \rbrace,
\label{22}
\end{eqnarray}

\begin{eqnarray}
\pi_{a (\lambda)} &=& i{\overline{\lambda}}_a, \qquad
\pi^a_{(\overline{\lambda})}= i\lambda^a,
\label{23}
\end{eqnarray}
one can conclude, that the system possesses the secon-class fermionic
constraints

\begin{eqnarray}
\Pi_{(\lambda)_a} &=& \pi_{(\lambda)_a} - i \overline{\lambda}_a, \qquad
\Pi^a_{(\overline{\lambda})} = \pi^a_{(\overline{\lambda})} -
i\overline{\lambda}^a,
\label{24}
\end{eqnarray}
since

\begin{equation}
\lbrace \Pi^a_{(\overline{\lambda})}, \Pi_{(\lambda)b}\rbrace = 2i\delta^a_b.
\label{25}
\end{equation}
Therefore, the quantization has to be done using the Dirac brackets defined
for any two functions $F$ and $G$

\begin{equation}
\lbrace F, G \rbrace^{\ast} = \lbrace F, G \rbrace -
\lbrace F, \Pi_a \rbrace \frac{1}{\lbrace \Pi_a, \Pi_b \rbrace }
\lbrace \Pi_b, G \rbrace.
\label{26}
\end{equation}
As a result, we obtain the following Dirac brackets for the canonical
variables

\begin{eqnarray}
\lbrace R, P \rbrace^{\ast} &=& 1, \qquad
\lbrace \lambda^a, {\overline{\lambda}}_b \rbrace^{\ast}=
\frac{i}{2}\delta^a_b.
\label{27}
\end{eqnarray}

The supercharges and the Hamiltonian form the following $n=4$ SUSY QM algebra
with respect to the introduced Dirac brackets

\begin{equation}
\lbrace {\bar Q}_a , Q^b \rbrace^{\ast} = -i \delta^b_a H_{0},  \quad
\lbrace {\cal F}_i , {\cal F}_j \rbrace^{\ast} = \epsilon_{ijk} {\cal F}_k ,
\label{28}
\end{equation}

\begin{equation}
 \lbrace {\cal F}_i , {\overline{Q}}_a \rbrace^{\ast} =
\frac{i}{2} {(\sigma_i)}^{c}_a {\overline{Q}}_c, \quad
\lbrace {\cal F}_i ,  Q^a \rbrace^{\ast} =
- \frac{i}{2} {(\sigma_i)}^{a}_c  Q^c.
\label{29}
\end{equation}
On the quantum level we replace the Dirac brackets by (anti)commutators
using the rule
\begin{equation}
i\lbrace , \rbrace^{\ast} = \lbrace , \rbrace.
\label{30}
\end{equation}
One obtains the non-zero commutation relations for the Dirac brackets

\begin{equation} \label{xx}
\lbrack R, P \rbrack = i, \qquad
\lbrace \lambda^a, {\overline{\lambda}}_b \rbrace = -\frac{1}{2} \delta^a_b.
\label{31}
\end{equation}
In the quantum theory the first-class constraints (\ref{17}-\ref{20})
associated
with the invariance of the action (\ref{14}) become conditions on the
wave function $\Psi$ of the Universe.
Therefore, any physically allowed states
must obey the quantum constraints

\begin{eqnarray}
H_{0}\Psi&=& 0, \qquad Q^a\Psi =0, \qquad \overline{Q}_a\Psi=0, \qquad
{\cal F}_i\Psi=0.
\label{32}
\end{eqnarray}
The quantum generators $H_{0}, Q^a, {\overline{Q}}_a$ and ${\cal F}_i$ form
a closed superalgebra of the $n=4$ supersymmetric quantum mechanics

\begin{eqnarray}
\lbrace {\overline{Q}}_a, Q^b \rbrace &=& H_{0}\delta_a^b, \qquad
\lbrack {\cal F}_i, {\cal F}_j \rbrack = i\epsilon_{ijk} {\cal F}_k, \qquad
\lbrack {\cal F}_i, {\overline{Q}}_a \rbrack =
-\frac{1}{2} (\sigma_i)_a^b {\overline{Q}}_b, \label{33}\\
\lbrack {\cal F}_i, Q^a \rbrack &=& \frac{1}{2} (\sigma_i)_b^a Q^b.
\nonumber
\end{eqnarray}
In the usual quantization the even canonical variables are replaced by
operators $R \to R, P \to -i\frac{\partial}{\partial R}$ and the odd variables
$\lambda$ and $\overline{\lambda}$ after quantization become
 anticommiting operators in accordance with (\ref{xx}).
In order to obtain the quantum expression for the Hamiltonian $H_0$ and for
the supercharges $Q^a$ and $\overline{Q}_a$ we must solve the operator ordering
ambiguity. Such ambiguities always arise when, as in our case, the operator
expression  contains the product of non-commutating operators $R$ and $P$
\cite{13}.
Technically it means the following: for the quantum supercharges we take the
same order as for the operator (\ref{18},\ref{19}). These calculations
are performed doing the integration with the measure $R^{1/2}dR$. With
this measure the conjugate momentum
 $P = -i\frac{\partial}{\partial R}$ is non-Hermitian with
$P^{\dagger} = R^{-1/2} P R^{1/2}$. However, the combination
$(R^{-1/2} P)^{\dagger} = P^{\dagger} R^{-1/2} = R^{-1/2} P$ is Hermitian and
$(R^{-1/2} P R^{-1/2} P)^{\dagger} = R^{-1/2} P R^{-1/2} P$ is also Hermitian.
So the anticommutation relation $\lbrace Q^a, Q^{\dagger}_b \rbrace =
\delta^a_b H_0$ fix all
additional terms and define the quantum Hamiltonian, but in this case the
operational expression $-\frac{{\mbox{\ae}}^2}{12}(R^{-1/2}PR^{-1/2} P)$
corresponds to the energy of the scale factor $R$ in the Hamiltonian
(\ref{33}). As we can see from the classical Hamiltonian
(\ref{33}), the energy of the scale factor is negative. This is due to the fact
that the particle-like fluctuations do not correspond to the scale factor
$R(t)$.

 \section{Conclusions}
On the basis of local $n=4$ supersymmetry the superfield action for the FRW
cosmological model was formulated. Due to the quantum supersymmetric
algebra (\ref{33}), the Wheeler-DeWitt equation, which is of the
second-order, can be replaced by the four supercharge operator equations
constituting its supersymmetric ``square root".

It would be very interesting to generalize the proposed $n=4$ superfield
model to the all Bianchi-type models \cite{8}, as well as, to consider
interaction with matter fields and analize the spontaneous breaking of the
$n=4$ local supersymmetry. We hope that for this more general supersymmetric
cosmological models than \cite{11}, we can find a normalizable wavefunction.
The details of this searchs will be given elsewhere.

 \noindent {\bf Acknowledgments.}
 We are grateful to  J. Socorro and O. Obreg\'on for their interest
in the work and useful comments. This research was supported in part by
CONACyT under the grants 2845-54E and 28454E. Work of A.P. was supported
in part by the RFBR under the
grant 99-02-18417 and the joint grant RFBR-DFG 99-02-04022,
and work of M.T. by RFBR 99-02-18417 and RFBR 01-02-06531.
One of us (J.J.R) would like to thank CONACyT for support
under Estancias Posdoctorales
en el Extranjero.
 
\end{document}